\documentclass{mn2e}

\usepackage{psfig}

\def\ltsima{$\; \buildrel < \over \sim \;$}
\def\lsim{\lower.5ex\hbox{\ltsima}}
\def\gtsima{$\; \buildrel > \over \sim \;$}
\def\gsim{\lower.5ex\hbox{\gtsima}}
\def\mes{M\'esz\'aros}

\begin{document}
\title[Long lasting engines in GRBs]{X-ray flares and the duration of
engine activity in gamma-ray bursts}

\author[Lazzati \& Perna] {Davide Lazzati and Rosalba Perna \\ JILA,
University of Colorado, Boulder, CO 80309-0440, USA {\tt e-mail:
lazzati@colorado.edu, rosalba@jilau1.colorado.edu}}

\maketitle

\begin{abstract}
The detection of bright X-ray flares superimposed on the regular
afterglow decay in {\em Swift} gamma-ray bursts has triggered theoretical
speculations on their origin. We study the temporal properties of
flares due to internal dissipation and external shock mechanisms. We
first show that at least a sizable fraction of the flares cannot be
related to external shock mechanisms, since external shock flares
evolve on much longer time scales than observed. We then study flares
from internal dissipation, showing that the temporal properties allow
us to distinguish the emission of slow early shells from that of late
faster shells. We show that, due to the rapid evolution of the
detected flares, it is most likely that the flares are produced by
relatively fast shells ejected by the central engine shortly before
they are observed. This implies that the central engine must be active
for, in some cases, as long as one day.  We finally discuss the
constraints and implications that this observation has on the
properties and physics of the inner engine, and we elaborate on
possible future observational tests on the flare sample to further
understand their origin and physics.
\end{abstract}
\begin{keywords}
gamma-ray: bursts ---  radiation mechanisms: non-thermal
\end{keywords}

\section{Introduction}

The external shock model predicts afterglows characterized by a
power-law decay in time, with the possible presence of breaks
connecting branches of different slope (\mes~\& Rees 1997). Such
changes are due to either geometrical properties of the fireball
(geometrical beaming, Rhoads 1999) or to spectral transitions
(e.g. from a cooling to a non-cooling electron population, Sari, Piran
\& Narayan 1998). In the pre-{\em{Swift}} era, most afterglows were
consistent with the simplest version of this model. There were,
however, notable exceptions, such as GRB~000301C (Masetti et
al. 2000), GRB~021004 (Lazzati et al. 2002), and GRB~030329 (Matheson
et al. 2003). All these bursts displayed optical variability in the
form of bumps or wiggles superimposed on the smooth power-law decay.

A number of explanations have been discussed in the literature to
account for the variations in the optical afterglow brightness. These
include inhomogeneities in the external density (Wang \& Loeb 2000;
Lazzati et al. 2002; Heyl \& Perna 2003), refreshed shocks due to the
collision of a late shell of plasma with the external shock material
(Rees \& \mes~1998), angular inhomogeneities in the fireball energy
distribution (Nakar, Piran \& Granot 2002) and gravitational lensing
(Loeb \& Perna 1998; Garnavich, Loeb \& Stanek 2000). Different
mechanisms to produce variability in afterglows can in principle be
distinguished through their temporal and spectral properties. In
practice, however, a consensus has not been yet reached due to the
lack of unambiguous observations.

More recently, {\em Swift} observations have revealed that flaring
activity is relatively common in the early phases of GRB afterglows,
sometimes extending for over a day (e.g., Chincarini 2006; O'Brien et
al. 2006). The temporal properties of the flares, their intensity, and
their spectra suggest an origin that is unrelated to the external
shock (Chincarini 2006; Falcone et al. 2006), at least for a fraction
of the bumps. Late episodes of ``prompt emission''\footnote{Here and
in the following we will call prompt emission all the radiation
produced by the outflow inside the external shock. Related mechanisms
include, but are not limited to, internal shocks, magnetic
dissipation, and comptonization of external photons.} can have two
different origin. One possibility is that the inner engine itself is
active for a time as long as the detection time of the X-ray
flare. Alternatively, the engine can be short-lived but produce,
together with the fast ejecta, a tail of slower material. Such slower
material can produce internal dissipation (and therefore prompt
emission) at late time. Slow shells ejected immediately after the fast
ejecta will not produce an external shock at the canonical external
shock radius since the ambient medium has been swept by the external
shock produced by the fast ejecta, which develops
earlier. Distinguishing between these two scenarios bears important
implications for the physics of the GRB engine.

In this paper we discuss the timescales of flares in the different
scenarios.  We show that at least a large fraction of X-ray flares are
due to a long lasting activity of the GRB engine, rather than to
external shock activity or to the emission of slow shells immediately
after the prompt emission ends. 

This paper is organized as follows: in Sect.~2 we compute external
shock timescales while in Sect.~3 we consider prompt emission flares
and in Sect. 4 we discuss our results and compare them to {\em Swift}
data. We summarize our findings in Sect.~5.

\section{External shock}

\begin{figure}
\psfig{file=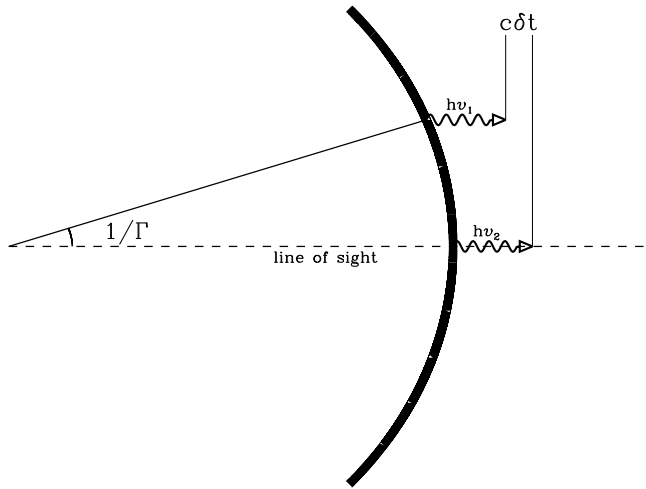,width=\columnwidth}
\caption{{Cartoon showing the geometry of the angular time scale.}
\label{fig:curva}}
\end{figure}

Consider flaring activity produced by a sudden brightening of the
external shock. We consider now a brightening that involves the whole
surface of the external shock. Even if the duration of the activity in
the comoving frame is negligible, the observer at infinity still
detects photons over a finite amount of time, due to the ``curvature
effect'' (see Fig.~\ref{fig:curva}). The shape of the observed pulse
is therefore the narrowest possible in time. A longer activity in the
comoving frame will produce a longer pulse, obtained by convolution of
the two functional shapes.

Let us define the quantity $E^\prime(\nu^\prime)$ to be the energy
released by the unit area of the fireball as a result of the flaring
activity, in the comoving frame. The flux received at infinity is
given by (see also Kumar \& Panaitescu 2000):
\begin{equation}
F_{\nu}(t) = \frac{1}{4\pi\,D^2}
\frac{E^\prime(\nu/\delta)\,\delta^2\,d\Sigma}{dt_{\rm{obs}}}\,,
\label{eq:fnu}
\end{equation}
where $D$ is the distance to the source,
$\delta\equiv[\Gamma(1-\beta\cos\theta)]^{-1}$ is the Doppler factor,
$\Sigma$ is the fireball surface and $t_{\rm{obs}}$ the time in the
observer frame. This is given by:
\begin{equation}
t_{\rm{obs}}=\int_0^R\frac{dr}{\beta_{\rm{sh}}\,c}
-\frac{R}{c}\cos\theta 
\simeq \frac{R}{c}\left[1-\cos\theta+\frac{1}
{2\Gamma^2_{\rm{sh}}(2\alpha+1)}\right]\,.
\label{eq:tobs}
\end{equation}

Equation~(\ref{eq:tobs}) holds for any self-similar dynamical evolution
where $\Gamma_{\rm{sh}}\propto{R}^{-\alpha}$, as long as
$\Gamma_{\rm{sh}}\gg1$. Note that we indicate with the subscript
$_{\rm{``sh''}}$ quantities (such as Lorentz factor and speed) of the
external shock, and without subscript the same quantities for the
material just behind the shock. It can be shown that
$\Gamma_{\rm{sh}}=\sqrt{2}\,\Gamma$ (e.g., Sari 1997).

Easy considerations show that the only non constant quantity in
Eq.(~\ref{eq:fnu}) is the Doppler factor, since
$d\Sigma/dt_{\rm{obs}}=2\pi\,R\,c$, and all the radiation is emitted at
constant radius $R$. If we assume that we are far from a break in the
afterglow spectrum, $E^\prime(\nu^\prime)\propto(\nu^\prime)^{-\eta}$
and the functional shape of $F_{\nu}(t)$ can be rewritten as:
\begin{equation}
F_{\nu}(t)\propto\delta^{(2+\eta)}\,.
\end{equation}

Solving Eq.~(\ref{eq:tobs}) for $\cos\theta$, and expressing time in
unit of the start time of the flare
$t_0=R/[2c\,\Gamma_{\rm{sh}}^2(2\alpha+1)]$, the flare profile can be
written as:
\begin{equation}
F_{\nu}(t)\propto\left(1+\frac{\tau-1}{4\alpha+2}\right)^{-(\eta+2)}\,,
\label{eq:fnu2}
\end{equation}
where $\tau=t/t_0$. 

\begin{figure}
\psfig{file=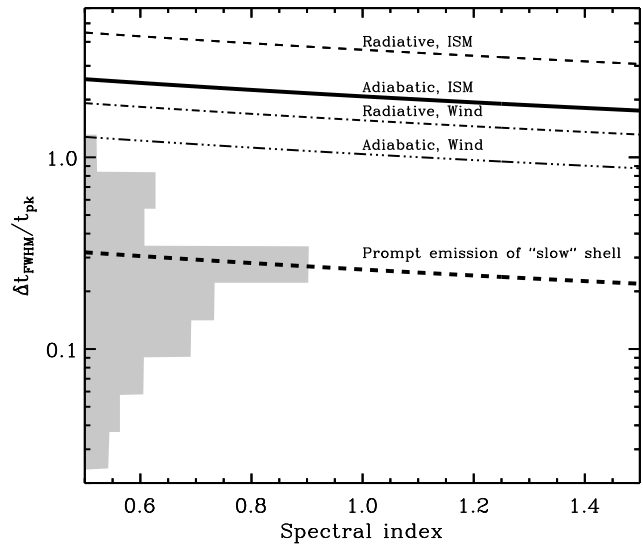,width=\columnwidth}
\caption{{Minimum time scale ratio for flares in different dynamical
settings as a function of the spectral index $\eta$. From top to
bottom, we show a radiative shock propagating into a uniform medium
($\alpha=3$), an adiabatic shock propagating into a uniform medium
($\alpha=3/2$), a radiative shock propagating into a wind profile
($\alpha=1$), an adiabatic shock propagating into a wind profile
($\alpha=1/2$), and the limit for internal dissipation from a slow
shell ejected during or immediately after the prompt phase. The shaded
area represents the observational distribution of $\Delta{t}/t$ from
Chincarini (2006), corrected for our $\Delta{t}$ definition.}
\label{fig:dtt}}
\end{figure}

Equation~\ref{eq:fnu2} can be used to provide a lower limit to an
observationally sound definition of the flare duration. Defining the
duration as the Full Width at Half Maximum (FWHM) of the pulse, and
the reference time as the time t$_{\rm pk}$ of the flare
maximum, we obtain:
\begin{equation}
\frac{\Delta{t}}{t} \equiv 
\frac{\Delta{t}_{\rm{FWHM}}}{t_{\rm{pk}}}
\ge \left(2^{\frac{1}{\eta+2}}-1\right)(4\alpha+2)\,.
\label{eq:dtt}
\end{equation}

Equation~(\ref{eq:dtt}) depends on both the dynamics and spectrum of
the external shock. A steeper spectrum and/or a more slowly evolving
fireball will cause a narrower flare. Note that in this case
(i.e. impulsive rebrightening of the shock), the maximum coincides
with the start time. Should the activity in the comoving frame last
for a finite amount of time, a rising phase would be observed. A more
detailed treatment, including the width of the shell of shocked
material would also provide a profile with a finite rising time.
Figure~\ref{fig:dtt} shows the time scale ratio for four typical
values of $\alpha$ as a function of the spectral index. The most
relevant case for X-ray flares is the adiabatic ISM case, with a
spectral slope $\eta\sim1.2$. We find for that specific case:
$\Delta{t}/t\gsim2$.

In an analogous context, Zhang et al. (2006) discussed the possible
production of flares by various mechanisms such as density bumps,
post-energy injection in the blast wave, and two-components or patchy
jets. In all cases, they concluded that the characteristics of the
flares were not generally consistent with what predicted by these
models (i.e. $\Delta t/t\sim 1$ and a slower-than observed flux
decline).  Here, we have derived detailed, quantitative, upper limits
on the value of $\Delta/t$.  Our constraints are somewhat tighter than
what previously derived by different authors in the context of the
external model for flares (e.g., Ioka, Kobayashi \& Zhang 2005) but
comparable to more recent results (Wu et al. 2006).  Our method
differs in two important ways from what done in previous work.  First,
we have computed the actual functional shape of the flare, and defined
rigorously the time reference we adopt (the peak) and the time width
(the FWHM). In previous works, these quantities were usually poorly
defined, and the results could only be approximate. Secondly, we
considered the difference between the Lorentz factor of the shock and
that of the shocked material. This factor by itself leads to flares
which are broader by a factor of two.

Chincarini (2006) presents a distribution of $\Delta{t}/t$ for a
sample of {\em Swift} X-ray flares that is defined as the ratio of the
Gaussian $\sigma$ over the peak time. The distribution is shown in
their Fig.~8 and shaded in Fig.~\ref{fig:dtt}. Since for a Gaussian
the $FWHM$ is equal to $2.35\sigma$, we can easily convert their
definition of $\Delta{t}$ into our definition. We conclude that the
distribution of $\Delta{t}/t$ of {\em Swift} X-ray flares is
characterized by $0.02\lsim\Delta{t}/t\lsim 1.3$ and has average
$\Delta{t}/t=0.3$. Considering Fig.~\ref{fig:dtt}, we conclude that
most, if not all, the detected rebrightenings are not related to
events taking place on the external shock, unless only a small portion
of the shock is involved in the rebrightening. Flares caused by a
small portion of the shock are possible, but are unlikely to be very
strong. Nakar \& Granot (2006) studied the effect of a blob of high
density interacting with the shock. They find that the rebrightening
is minor, and characterized by a very slow rise. Alternatively, a
large rebrightening on a small timescale can be observed if a narrow
opening angle shell refreshes a small part of the external shock
(Granot, Nakar \& Piran 2003). This is not likely to be the case for
the {\em Swift} X-ray flares, since most of them are observed before
the jet break time (Chincarini 2006, O'Brien 2006) and the opening
angle of the jet grows with time (Lazzati \& Begelman 2005; Morsony,
Lazzati \& Begelman 2006) rather than decrease.

\section{Prompt emission}

Since the largest majority of flares is unlikely to be produced within
the external shock, it must be produced during the initial phase of
internal dissipation. In the following we therefore discuss flaring
activity in the context of the prompt emission.  In this case one can
envisage two possible scenarios for their production: {\em (a)} shells
that are produced during the GRB phase but that dissipate at much
later times; {\em (b)} shells produced at late times by a long-lived
engine and dissipating on a timescale comparable to that of the engine
duration.  Zhang et al. (2006) also concluded that the flares have an
internal origin, but directly assumed that the engine must have been
long-lived. In this section we try to discriminate between the
scenarios {\em (a)} and {\em (b)}. Being able to discriminate between
these two scenarios bears important implications for our understanding
of the physics of the inner GRB engine (see more discussion on this in
\S4). In the following, we discuss the two scenarios above, provide
diagnostics for each of them, and show that the current {\em Swift}
data is already able to discriminate between the two flare-production
mechanisms.

{\em (a) Flaring activity due to dissipation within a
freely expanding flow, released during the prompt GRB phase or shortly
afterwards.} This case has two important differences with respect to the
external shock case studied above. First, there is no propagating
shock, and therefore there is no distinction between $\Gamma$ and
$\Gamma_{\rm{sh}}$. Second, the flare is produced by material that has
been coasting at constant $\Gamma$, rather than by material that has
been slowing down. These two differences change considerably the
equations above.

Without repeating the derivation of Sect.~2, it can be easily seen
that the result for the freely expanding flow can be obtained by
substituting $4\alpha+2$ with $2\alpha+1$ and adopting $\alpha=0$ in
Eq.~\ref{eq:dtt}.  This yields:
\begin{equation}
\frac{\Delta{t}_{\rm{FWHM}}}{t_{\rm{pk}}} \gsim
2^{\frac{1}{2+\eta}}-1\,,
\label{eq:dtt2}
\end{equation}
which, for the typical case of $\eta\sim1.2$ yields
$\Delta{t}/t\gsim0.25$ (see Wu et al. 2006 for a similar computation
restricted to the case of internal shocks). This number is tantalizing
similar to the average width of {\em Swift} X-ray flares (Chincarini
2006; Fig.~\ref{fig:dtt}). Yet, there is a sizable number of events
for which the condition is violated by almost an order of
magnitude. In addition, should the condition be barely satisfied, the
decay part of the flare would be dominated by large angle emission
(Kumar \& Panaitescu 2000; Liang et al. 2006). This seems not to be
the case, at least in some of the flares (Guetta et al. 2006). Note
however that the large angle emission should not dominate any flare
with a duration significantly larger than the one given in
Eq.(~\ref{eq:dtt2}).

{\em (b) Shell that is ejected from the central engine after a time
$t_{\rm{ej}}$ sizably larger than the prompt emission timescale
$T_{90}$.} If radiation is released at radius $R$, it will be observed
at a time (cfr. Eq.~\ref{eq:tobs})
\begin{equation}
t_{\rm{obs}} =
t_{\rm{ej}}+\frac{R}{c}\left(1-\cos\theta+\frac{1}{2\Gamma^2}\right)\,,
\end{equation}
which produces a shortening of the observed variability timescale
\begin{equation}
\frac{\Delta{t}}{t} \gsim \left(2^{\frac{1}{2+\eta}}-1\right) \,
\left(1-\frac{t_{\rm{ej}}}{t_{\rm{pk}}}\right)\,.
\label{eq:dtts}
\end{equation}

Equation~(\ref{eq:dtts}) allows for any arbitrarily small time scale, as
long as the shell that produces radiation is ejected at a time comparable to
the detection time. Therefore, the detection of fast flares implies 
that the inner engine is active for a time much longer than the
canonical prompt phase ($T_{90}$).

The observation of a flare that peaks at time $t_{\rm{pk}}$ and is
characterized by a width $\Delta{t}$ allows us to constrain the time
at which it was ejected from the central engine:
\begin{equation}
1-\frac{1}{2^{\frac{1}{2+\eta}}-1}\frac{\Delta{t}}{t}
\lsim\frac{t_{\rm{ej}}}{t_{\rm{pk}}}\le1 \,.
\label{eq:wow}
\end{equation}

\section{Discussion}

Equations~(\ref{eq:dtt2}) and~(\ref{eq:dtts}) are general and hold for
any dissipation and radiation mechanism that take place in the whole
jet. Locally beamed emission mechanisms can produce fast variability
(see, e.g., Lyutikov \& Blandford 2004), but the peak luminosity of
the flares would drop faster than observed as a function of
$t_{\rm{pk}}$. Consider now the flares observed by {\em Swift} (for a
recent review, see {\tt
http://www.swift.ac.uk/rs06/Burrows.pdf}). They can be observed up to
$\sim10^5$~s after the GRB onset and are characterized by a FWHM time
$\delta{t}\sim0.3\,t$. Applying this observational constraint implies
that at least in about half of the cases the ejection time of the
material from the central engine is comparable to the time at which
the flare is observed. In other words, GRB engines are active well
beyond the observed $T_{90}$ of the prompt emission, in some cases for
up to $\sim10^{5}$~s or a day timescale.

This provides an important constraint to the properties of the inner
engine of long duration GRBs. More can be obtained by further analysis
of the properties of the flares. Flares are observed over a wide range
of times, spanning about three orders of magnitude (from $100$ to
$10^5$~s after the end of the prompt emission). The distribution of
$\delta{t}/t$ is however much narrower, spanning approximately only
one order of magnitude. This implies a correlation between the
ejection time and the duration of the ejection episode. The longer is
the time after the engine turns on, the longer is the ejection
episode. Interestingly, an analogous property was noted in GRBs with
multiple prompt emission episodes (Ramirez-Ruiz \& Merloni
2001). During the prompt phase, the duration of an emission episode is
strongly correlated with the length of the quiescent time that
precedes it. In the case of flares we can conclude that the
correlation is with the total time since the engine onset; however,
this does not imply the absence of a correlation with the length of
the quiescent time.

Our results bear important implications for an understanding of the
physical processes governing the GRB inner engine, which provides the
ultimate source of energy to power GRBs. The GRB ``inner engine'' is
believed to be a hyperaccreting accretion disk. In the collapsar model
(Paczynsky 1998; MacFadyen \& Woosley 1999), this is formed by the
fallback material from the collapsing envelope of the massive star,
and the timescale for the duration of the accretion episode is set by
the dynamical timescale of the collapsing envelope itself, which is on
the order of several tens of seconds.  On the other hand, in the
binary merger model (e.g. neutron star-neutron star or neutron
star-black hole; Eichler et al. 1989), the accretion material is
provided by the debris of the tidally disrupted neutron star (or white
dwarf). In this case, the timescale over which the available material
is accreted is set by the viscous timescale of the disk, which is a
fraction of a second.

The accretion timescale in the collapsar model easily accounts for the
duration of the prompt emission for the class of long bursts, while
the accretion timescale in the binary merger model naturally yields
the required timescales for the prompt emission of short bursts.
Several types of observations in the last few years have indeed
provided a strong support for the association between long GRBs and
the collapse of a massive star (Stanek et al. 2003; Hjorth et
al. 2003), while several pieces of evidence are gradually mounting
toward the association of compact-object mergers and short bursts
(Fox et al. 2005; Bloom et al. 2006; Nakar, Gal-Yam \& Fox 2006).

In the scenarios above, a relativistic outflow is ejected from the
central engine, and radiation is produced as the bulk kinetic energy
is dissipated and radiated at a given radius. There is no a-priori
constraint on the distribution of the Lorentz factor in the outflow.
If the Lorentz factor of the ejecta should decrease with the ejection
time, late time internal activity could be seen without the need for a
long-living engine. We have shown that this is not the case, at least
in a sizable fraction of the flares.  This result makes the studies of
possible mechanisms of reactivation of the GRB engine especially
timely.  King et al. (2005) suggested that fragmentation of the
collapsing star could explain a (single) flare in the case of long
bursts, while Dai et al. (2006) proposed a new mechanism of binary
merger that can incorporate the presence of a flare for short bursts.
Perna, Armitage \& Zhang (2006) noted that the fact that flares are
observed in both long and short bursts, and with similar
characteristics in both cases, is suggestive that the place of
formation be in what is common between the two classes of bursts,
i.e. the accretion disk. A disk that fragments as a result of
gravitational instabilities in its outer parts, as supported by
studies of hyperaccreting disks (Di Matteo, Perna \& Narayan 2002;
Chen \& Beloborodov 2006), can account for flares with similar
properties in both long and short bursts, as well as accounting for
the observed correlation between the flare duration and its arrival
time (see also Piro \& Pfahl 2006).  Other disk-based models involve
magnetic instabilities (Proga \& Zhang 2006; Giannios 2006). Staff,
Ouyed \& Bagchi (2006) proposed instead a mechanism based on state
transitions in the quark-nova scenario, while the role of a magnetar
was discussed by Gao \& Fan (2006) and Cea (2006).

\subsection{Refreshed shocks}

If an ``internal'' flare, i.e., a flare due to internal dissipation,
is observed in a GRB, it must be followed at later time by an external
flare. The material responsible for the internal flare keeps expanding
at constant speed, and must eventually catch up with the external
shock, which decelerates continuously. It may however be difficult to
observe the second flare, since it may be weak and dispersed over
several orders of magnitudes in time. If, however, the second flare
can be singled out and firmly associated to an external flare, it
would provide additional important information for our understanding
of the dynamics of the system.

Consider a flare system made by an internal flare beginning at time
$t_{\rm{int}}$ followed by a second, broader, flare starting at time
$t_{\rm{ext}}$, where $t_{\rm{ext}}\gg{t}_{\rm{int}}$. The external
flare time can tell us the Lorentz factor of the material producing
the internal flare:
\begin{equation}
\Gamma_{\rm{ej}} \simeq \left(\frac{3E}{8\pi\,n\,m_p\,c^5}\right)^{1/8}
\,t_{\rm{ext}}^{-3/8}\,,
\label{eq:gamma}
\end{equation}
where $E$ is the isotropic equivalent kinetic energy of the fireball
integrated up to the time of ejection of the late shell, $n$ the
numeric density of the external medium, and $m_p$ the proton mass. We
have assumed a uniform external medium. Equations can be easily
generalized to an arbitrary density profile, but will be simple only
in the case of adiabatic evolution. Equation~(\ref{eq:gamma}) depends
only on measurable quantities.  An example of a solvable system is the
flare system of GRB~050502B. Falcone et al. (2006) were able to
identify an initial strong flare and a possible refreshed shock flare
at later times. They were able to measure the bulk Lorentz factor of
the late engine ejecta to be $\Gamma_{\rm{ej}}\lsim20$, ejected from
the inner engine approximately $300$ second after the onset. By itself
this observation only tells us that the late ejections are likely to
have a smaller Lorentz factor than the early ones. Solving a large
numbers of flare systems in different GRBs would however provide us
with an ensemble of late ejections that would allow us to constrain
both the properties of the engine and of the dissipation/radiation
mechanism. The knowledge of the bulk Lorentz factor of the late shells
would provide us with a sample of flares for which the comoving
spectra are known, ridding the prompt emission models of one of the
unknown parameters: the Lorentz factor.

\section{Summary}

We have analyzed the timescale of flares produced by several
mechanisms both due to internal dissipation and external shock
phenomena. We have shown that all prominent external shock flares
occurring before the jet break must have a long time scale
$\delta{t}\gsim{t}$. We also showed that internal dissipation taking
place in slow shells ejected immediately after the end of the prompt
phase must satisfy a similar constraint: $\delta{t}\gsim{0.25t}$. The
only mechanism capable of producing flares with a short time scale as
observed by {\em Swift} is identified with late activity of the inner
engine. As a consequence, the detection of X-ray flares at times as
long as $10^5$ seconds after the GRB onset implies that the inner
engine is active for at least $10^5$~s. This has important
implications for the fueling of the engine and the ejection mechanism.

\section*{Acknowledgements}

We thank Dafne Guetta and Luigi Piro for useful conversations.  This
work was supported by NASA Astrophysical Theory Grant NNG06GI06G (DL),
NSF grant AST-0307502 (DL) and AST 0507571 (DL \& RP), and {\em Swift}
Guest Investigator Program NNX06AB69G (DL) and NNG05GH55G (DL \& RP).

\end{document}